\documentclass[%
 reprint,
 amsmath,amssymb,
 aps,
 prb,
]{revtex4-2}

\newcommand*\dif{\mathop{}\!\mathrm{d}}

\renewcommand{\Re}{\operatorname{Re}}
\renewcommand{\Im}{\operatorname{Im}}

\DeclareMathOperator{\Tr}{Tr}
\usepackage{graphicx}
\usepackage{dcolumn}
\usepackage{xcolor}
\newcommand*\VF[1]{\mathbf{#1}}
\usepackage{bm}
\begin{document}
\preprint{APS/123-QED}
\title{Nonreciprocal radiative heat transfer between two planar bodies}
\author{Lingling Fan$^1$, Yu Guo$^1$, Georgia T. Papadakis$^1$, Bo Zhao$^1$, Zhexin Zhao$^1$,\\ Siddharth Buddhiraju$^1$, Meir Orenstein$^2$, and Shanhui Fan$^1$}\email{shanhui@stanford.edu}

\affiliation{$^1$Department of Electrical Engineering, Ginzton Laboratory, Stanford University, Stanford, CA 94305, USA\\$^2$Department of Electrical Engineering, Technion-Israel Institute of Technology, 32000 Haifa, Israel
}
\date{\today}
\begin{abstract}
{We develop an analytical framework for nonreciprocal radiative heat transfer in two-body planar systems. Based on our formalism, we identify effects that are uniquely nonreciprocal in near-field heat transfer in planar systems. We further introduce a general thermodynamic constraint that is applicable for both reciprocal and nonreciprocal planar systems, in agreement with the second law of thermodynamics. We numerically demonstrate our findings in an example system consisting of magneto-optical materials. Our formalism applies to both near- and far-field regimes, opening opportunities for exploiting nonreciprocity in two-body radiative heat transfer systems.}
\end{abstract}
 \maketitle
\section{\label{sec:level1}Introduction}Understanding radiative heat transfer [\onlinecite{PhysRevB43303_polder, 1989psr4bookFDT, Mulet_thermal_near}] is essential in many applications ranging from radiative cooling [\onlinecite{nl301708e_Lipson_cool, Raman2014, PhysRevB_kaifeng_cool}] and thermal diodes  [\onlinecite{PhysRevLett.104.154301}] to thermal transistors [\onlinecite{doi:10.1063/1.2191730,RevModPhys.84.1045, PhysRevLett_gate_euro, PhysRevApplied.11.024004}] and thermophotovoltaic systems [\onlinecite{Celanovic04,TPV_Joulain,Simovski13, umich_nn_TPV,Omair15356}]. The majority of works investigating radiative heat transfer consider materials that satisfy Lorentz reciprocity [\onlinecite{GratingRotate, yuguoSuperPlankian, PhysRevBIllicGraphene, PhysRevB_Kruger, Guo_hyper,PhysRevLett_ODM_effectiveness, PhysRevBbozhaoHyper,PhysRevLett_SiC_period,Miller2DNL,  doi:10.1063/1.5049471, acsphotonics2D_harbin,PhysRevApplied_MO_2d,PhysRevX_2019_HSOM, PhysRevB_ZMZhang_periodicAnisotro}]. On the other hand, it is known that breaking the constraint of reciprocity is necessary in order to reach the thermodynamic limit of thermal radiation harvesting [\onlinecite{doi:10.1063/1.328187, Ries1983,  doi:10.1021/nl3034784}]. Therefore, in recent years there has been emerging interest in exploring radiative heat transfer with nonreciprocal materials. Examples include the design of photonic structures for complete violation of Kirchhoff's law in far-field thermal radiation [\onlinecite{PhysRevB_linxiao_farField_vio}], as well as the proposal for the thermal Hall effect [\onlinecite{PhysRevLett.116.084301}], and persistent heat current in equilibrium in near-field heat transfer [\onlinecite{PhysRevLett_linxiao_flow}]. 
\par{For near-field heat transfer, the most studied geometry is that between two planar bodies, separated by a vacuum gap (Fig.~\ref{fig:fig1_yes}), whereas previous works on nonreciprocal near-field heat transfer focus on non-planar geometries [\onlinecite{PhysRevLett_linxiao_flow, PhysRevB_Linxiao_general_theory,guo2019relation}]. There has not been a systematic study as to how nonreciprocity manifests itself in planar systems. Ref. \onlinecite{PhysRevB_Cuevas_2015} considered near-field heat transfer between two planar structures incorporating magneto-optical materials, but did not specifically address whether the observed effect is a manifestation of nonreciprocity. } 
\par{In this paper, we provide a systematic treatment of near-field heat transfer between two planar bodies. We address how the second law of thermodynamics and reciprocity constrain heat transfer in these systems. Specifically, we highlight signatures that are uniquely nonreciprocal, i.e., they cannot occur in reciprocal systems. Our work clarifies the role of nonreciprocity in one of the most commonly studied near-field heat transfer geometries, and may prove useful for the design of nonreciprocal heat transfer devices. }
\par{The paper is organized as follows: In Section~\ref{setup_sec}, we provide a derivation of the formalism for computing near-field heat transfer between planar bodies. This formalism is similar to the standard treatment of radiative heat transfer with fluctuational electrodynamics [\onlinecite{PhysRevB43303_polder}], however, here we place special emphasis to ensure that it is applicable to both nonreciprocal and reciprocal systems. Using this formalism, in Section~\ref{Constraint_sec} and \ref{Sec:rec_const}, we address how the second law of thermodynamics and reciprocity constrain near-field heat transfer in planar geometries. We highlight the uniquely nonreciprocal aspects of near-field heat transfer that cannot exist in reciprocal systems. In Section~\ref{numerical_sec}, we provide numerical demonstration of the theoretical predictions in Section~\ref{Constraint_sec} and \ref{Sec:rec_const}. We conclude in Section~\ref{conclude_sec}.  }\section{\label{setup_sec} Theoretical formalism}
 \begin{figure}
 \includegraphics[scale = 0.3]{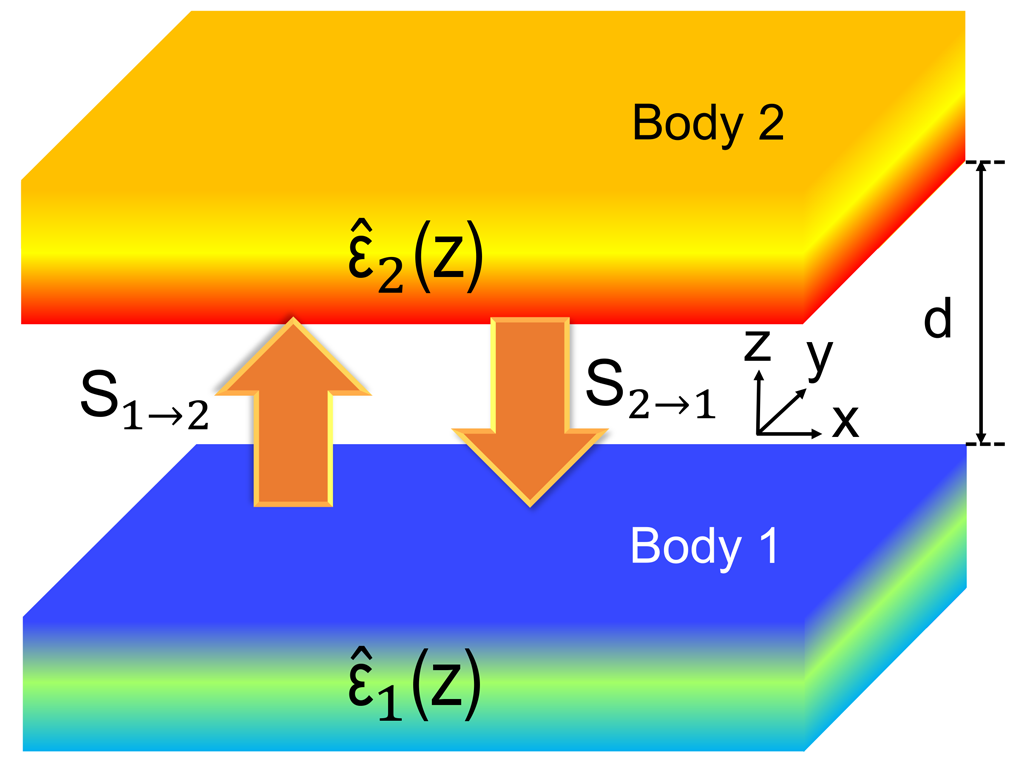} 
\caption{\label{fig:fig1_yes} Schematic of the geometry under consideration: Two semi-infinite planar slabs are separated by a vacuum gap of size $d$. Each slab has an in-plane ($x-y$) homogeneous dielectric permittivity, which can be inhomogeneous in the $z$-direction, represented with $\hat\epsilon_1(z)$, $\hat\epsilon_2(z)$. $S_{1\to 2}$ and $S_{2\to 1}$ represent the radiative heat flux density from body 1 to 2 and body 2 to 1, respectively. }
\label{fig:setup}
\end{figure}
\par{Throughout the paper, we consider the general heat transfer setting as shown in Fig.~\ref{fig:fig1_yes}. Body 1 and body 2 are semi-infinite, separated by vacuum with a gap size of $d$, and maintained at temperatures $T_1$ and $T_2$, respectively. Body 1 and 2 are described by a dielectric permittivity distribution $\hat{\epsilon}_1(z)$ and $\hat{\epsilon}_2(z)$, respectively, which are uniform in the in-plane directions, but can be nonuniform along the $z$-direction. In general, the tensors $\hat\epsilon_1(z)$, $\hat\epsilon_2(z)$ are $3\times 3$ permittivity tensors. From the fluctuation dissipation theorem [\onlinecite{laudau_book}], the strength of the fluctuating current sources that generate thermal radiation is proportional to the imaginary part of the permittivity tensor, $\Im \hat\epsilon\equiv\frac{1}{2j} (\hat\epsilon - \hat\epsilon^{\dagger})$.} 
\par{First, we compute the heat flux density from body 1 to body 2, ${S}_{1\to2}(\VF{r}_{\parallel},z,t)$, which is defined as the absorption of body 2 from the electric and magnetic field vectors $\VF{E}(\VF{r}_{\parallel},z,t)$ and $\VF{H}(\VF{r}_{\parallel},z,t)$ generated by the fluctuating current sources in body 1, and $\VF{r}_{\parallel}$, $z$, and $t$, are, respectively, the in-plane spatial dimension, distance along the vertical axis, and time. In the vacuum region between the two planar bodies in Fig. \ref{fig:fig1_yes}, the Poynting vector along the $z$ direction is given by:
\begin{align}
    S_{1\to2}(\VF{r}_{\parallel},z,t) = \hat{\VF{z}}\cdot\left<\VF{E}(\VF{r}_{\parallel},z,t)\times\VF{H}(\VF{r}_{\parallel},z,t)\right>
    \label{Eq: PoyntingFlux},
  \end{align}where $\left<\cdots\right>$ represents an ensemble average. Throughout the paper we adopt the following Fourier transformation conventions in time and space, respectively:
\begin{align}
    A(t) &= \Re\int_0^{\infty}{\dif{\omega}}A(\omega)e^{j\omega t},\label{time_fourier}\\
    A(\VF{r}_{\parallel}) &= \int\frac{\dif{\VF{k}_{\parallel}}}{(2\pi)^2}A(\VF{k}_{\parallel})e^{-j\VF{k}_{\parallel}\cdot\VF{r}_{\parallel}}\label{space_fourier}.
\end{align}$S_{1\to2}(\VF{r}_{\parallel}, z, t)$ is independent of $\VF{r}_{\parallel}$ by translational symmetry, and independent of $t$ since the thermal process is a stationary random process [\onlinecite{laudau_book_part1}, \onlinecite{Rytov_book}]. Therefore, we have
\begin{align}
  &~~~~\left<\VF{E}(\VF{k}_{\parallel},z,\omega)\times\VF{H}^*(\VF{k}_{\parallel}',z,\omega')\right>\nonumber\\ &=  \left<\VF{E}(\VF{k}_{\parallel},z,\omega)\times\VF{H}^*(\VF{k}_{\parallel},z,\omega)\right>\delta(\VF{k}_{\parallel}-\VF{k}_{\parallel}')\delta(\omega-\omega'),\label{Ehcorrelate}
\end{align}where $\VF{k}_{\parallel}$, $\VF{k}_{\parallel}'$ are the in-plane wavevector, and ${\omega}$, ${\omega}'$ are the frequency. With these notations from Eqs.~\ref{Eq: PoyntingFlux}-\ref{Ehcorrelate}, we obtain, 
\begin{align}
 &~~~~S_{1\to2}(\VF{r}_{\parallel}, z, t)\nonumber\\& = \int_0^{\infty}\dif{\omega}\int\frac{\dif{\VF{k}_{\parallel}}}{(2\pi)^4}\frac{1}{2}\Re\hat{\VF{z}}\cdot\left<\VF{E}(\VF{k}_{\parallel},z,\omega)\times\VF{H}^*(\VF{k}_{\parallel},z,\omega)\right>.
    \label{Eq:poyntingflux_fourier}
\end{align}Below we treat each $\VF{k}_{\parallel},~\omega$ components separately at a fixed $z$ in the vacuum gap, and suppress the arguments of $\VF{k}_{\parallel},~z$, and $\omega$.}
\par{We aim to express the near-field heat transfer between the two planar bodies, i.e., Eq.~\ref{Eq:poyntingflux_fourier}, in terms of the reflectivity matrix of each body. For this, we start by considering the electric field $\VF{E}_1$ as generated by body 1 in the absence of body 2 (Fig.~\ref{fig:fig2_yes}a).
Note that in the Appendix, we provide an analytical derivation of the field emission relations shown below. For propagating waves, defined to have an in-plane wavevector ${k}_{\parallel}<\frac{\omega}{c}$, where $c$ is the speed of light in vacuum, immediately near the surface of body 1, the field emission is expressed as:
\begin{align}
  \left<\VF{E}_1\VF{E}_1^{\dagger}\right>  &=(2\pi)^2\frac{Z}{\pi}\Theta(\omega,T_1)\left[\hat{I}-{\hat{R}}_1{\hat{R}}_1^{\dagger}\right],
  \label{eq:EEcorre_prop1}
  \end{align}whereas for evanescent waves with ${k}_{\parallel}>\frac{\omega}{c}$, we have:
\begin{align}
\left<\VF{E}_1\VF{E}_1^{\dagger}\right>   &=(2\pi)^2\frac{Z}{\pi}\Theta(\omega,T_1)\left[{\hat{R}}_1-{\hat{R}}_1^{\dagger}\right].
  \label{eq:EEcorre_evan1}
  \end{align}The parameter $Z$ in Eqs. \ref{eq:EEcorre_prop1}, \ref{eq:EEcorre_evan1} represents the wave impedance. We separate $Z$ into $Z_{\rm s}$ and $Z_{\rm p}$, i.e., $Z = {\rm{diag}}(Z_{\rm s}, Z_{\rm p})$, pertaining to $s$- and $p$-polarized waves, defined to have the electric field or magnetic field parallel to the material interfaces, respectively.  
The function $\Theta(\omega,T)$ is the mean energy of photons per frequency $\omega$ at temperature $T$, defined as $\Theta(\omega,T)=\hbar\omega\left[\frac{1}{2}+\frac{1}{{\rm{exp}}(\hbar\omega/k_{\rm B}T)-1}\right]$. Here, $\hbar$ is the reduced Planck constant, $k_{\rm B}$ is the Boltzmann constant. The reflectivity matrix of the $i$-th body $\hat{{R}}_{\rm i}$ is given by:
    \begin{align}
    \hat{{R}}_{\rm i} = \left[\begin{array}{cc}
 {R}_{\rm i}^{\rm ss} &  {R}_{\rm i}^{\rm sp}\\
    {R}_{\rm i}^{\rm ps} &   {R}_{\rm i}^{\rm pp}
    \end{array}\right],
    \label{reflec_mat}
\end{align}where $ {R}_{\rm i}^{\rm \sigma\mu}$ represents the reflection coefficient for light incident from vacuum into the $i$-th body, with a $\mu$-polarized incident wave and a $\sigma$-polarized outgoing wave. }
\begin{figure}[b]
\includegraphics[scale = 0.3]{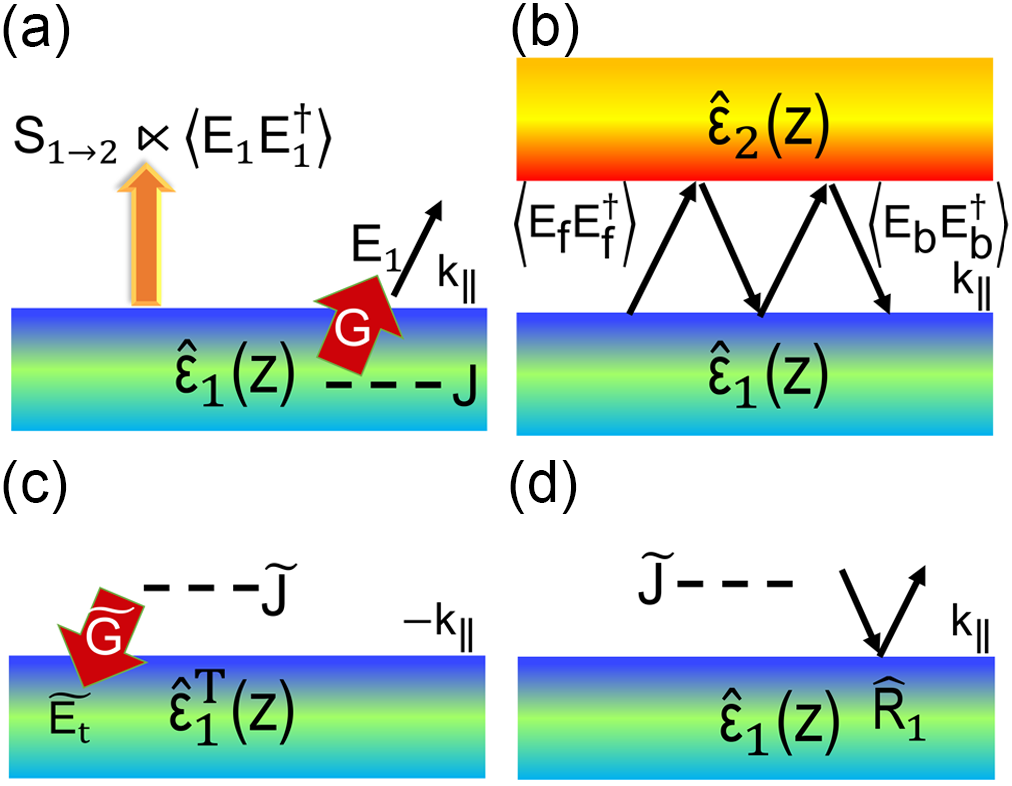} 
\caption{Theoretical framework for computing the heat transfer in a two-body planar system using the generalized reciprocal relation. (a) Thermal emission of body $1$. Poynting vector from body 1 to 2,  $S_{1\to2}(\omega,\VF{k}_{\parallel})$ is proportional to the electric field emission $<\VF{E}_1 \VF{E}_1^{\dagger}>$ near the source surface. (b) Heat exchange between bodies 1 and 2 via multiple scattering process. (c) and (d) illustrate the method to compute the field emission from a planar body using the generalized reciprocity relation. (c) The emission of body 1 can be equivalently computed by evaluating the absorption of the complementary body in the presence of the external source. (d) The absorption of the complementary body as shown in (c) can be computed in terms of its reflection coefficients.}
\label{fig:fig2_yes}
\end{figure}
\par{In the presence of body 2, the electric field between the two bodies can be decomposed into forward and backward components (Fig.~\ref{fig:fig2_yes}b), i.e., $\VF{E} =  \VF{E}_{\rm f} +  \VF{E}_{\rm b} $, which can be related to $\VF{E}_1$ as:  \begin{align}\VF{E}_{\rm f} &= \VF{E}_1 + \hat{R}_1 \hat{R}_2e^{-j2k_{\rm z} d}\VF{E}_1 + (\hat{R}_1 \hat{R}_2e^{-j2k_{\rm z} d})^2\VF{E}_1 +\cdots\nonumber\\ &= {{\hat{{D}}}}_{12}\VF{E}_1\\ \VF{E}_{\rm b}   &= \hat{R}_2{{\hat{{D}}}}_{12}e^{-j2k_{\rm z} d} \VF{E}_1 = {{\hat{{D}}}}_{21}\hat{R}_2e^{-j2k_{\rm z} d}\VF{E}_1,\end{align} where ${{\hat{{D}}}}_{\rm mn}=[\hat{I} - \hat{R}_{\rm m}\hat{R}_{\rm n}e^{-j2k_{\rm z}d}]^{-1}$ represents multiple reflections between the two bodies. In the alternative expression of $\VF{E}_{\rm b}$, we have used the identity,
\begin{align}
    \hat{R}_2\hat{D}_{12} = \hat{D}_{21}\hat{R}_2
    \label{Eq:Multiple_scattering}.
\end{align} By representing electric and magnetic fields with $\VF{E}_{\rm f}$ and $\VF{E}_{\rm b}$ components, the Poynting vector in Eq.~\ref{Eq:poyntingflux_fourier} can then be written as:
\begin{align}
    \frac{1}{2}\Re\{\hat{\VF{z}}\cdot\left<\VF{E}\times\VF{H}^*\right> \}
    &=\frac{1}{2} \Re\Tr\left[(\VF{E}_{\rm f} +\VF{E}_{\rm b})\left[{Z^{-1}}(\VF{E}_{\rm f} -\VF{E}_{\rm b})\right]^{\dagger}\right]
     \label{eq:totalEHzcorr}.
\end{align}which holds for propagating as well as evanescent modes, and $\VF{E}_{\rm f}$ and $\VF{E}_{\rm b}$ denote the forward and backward field components on the $s$ and $p$ polarization basis. Note that the basis vector for the $p$-polarization has opposite signs for forward and backward waves.
For propagating waves, Eq. \ref{eq:totalEHzcorr} becomes: 
\begin{align}
  &~~~~\frac{1}{2}\Re\{\hat{\VF{z}}\cdot\left<\VF{E}\times\VF{H}^*\right> \}\nonumber\\&=  \Tr\left[\frac{1}{2Z}(\hat{I} - \hat{R}_2^{\dagger}  \hat{R}_2) \hat{{D}}_{12}\left<\VF{E}_1\VF{E}_1^{\dagger}\right>\hat{{D}}_{12}^{\dagger}\right]
    \label{eq:scat_prop1},
\end{align}\noindent{whereas for evanescent waves, Eq. \ref{eq:totalEHzcorr}  becomes: 
\begin{align}
   &~~~~ \frac{1}{2}\Re\{\hat{\VF{z}}\cdot\left<\VF{E}\times\VF{H}^*\right> \}
   \nonumber\\ &= \Tr\left[\frac{1}{2Z}(\hat{R}_2^{\dagger}  -\hat{R}_2) {\hat{D}}_{12}\left<\VF{E}_1{\VF{E}}_1^{\dagger}\right>{\hat{D}}_{12}^{\dagger}e^{-2\alpha d}\right]
     \label{eq:scat_evan1}.
\end{align}In the latter case, the field components $\VF{E}_{\rm f}$ and $\VF{E}_{\rm b}$ denote exponential decay and growth, respectively, along the $z^+-$direction, where $k_{\rm z} = -j\alpha$, with $\alpha$ being a real number. }}
\par{Plugging Eq.~\ref{eq:EEcorre_prop1} and Eq.~\ref{eq:EEcorre_evan1} into Eq.~\ref{eq:scat_prop1} and Eq.~\ref{eq:scat_evan1}, respectively, and by recalling Eq.~\ref{Eq:poyntingflux_fourier}, we obtain the total heat flux density from body 1 to body 2: \begin{align}
     S_{1\to2}=\int_0^{\infty}\frac{\dif{\omega}}{2\pi}\int\frac{\dif{\VF{k}_{\parallel}}}{(2\pi)^2}{\Theta(\omega,T_1)}S_{1\to2}(\VF{k}_{\parallel}, \omega),\label{S12_integral_first}
\end{align}where $S_{1\to2}(\VF{k}_{\parallel}, \omega)$ is given by:
\begin{align} &~~~~S_{1\to2}(\VF{k}_{\parallel}, \omega)\nonumber\\ &=\Tr
\{ \left[\hat{I} - \hat{R}_2^{\dagger}(\VF{k}_{\parallel}, \omega) \hat{R}_2(\VF{k}_{\parallel}, \omega)\right]{\hat{D}_{12}}(\VF{k}_{\parallel}, \omega)\nonumber\\&~~~~\left[\hat{I} -{\hat{R}}_1(\VF{k}_{\parallel}, \omega){\hat{R}}_1^{\dagger}(\VF{k}_{\parallel}, \omega)\right] \hat{D}_{12}^{\dagger}(\VF{k}_{\parallel}, \omega)\}, \label{eq:wideeq1to2prop}\end{align} \noindent{for propagating waves, and}
 \begin{align}
 &~~~~S_{1\to2}(\VF{k}_{\parallel}, \omega)\nonumber\\
& =\Tr\{\left[\hat{R}_2^{\dagger}(\VF{k}_{\parallel}, \omega)-\hat{R}_2(\VF{k}_{\parallel}, \omega)\right] \hat{D}_{12}(\VF{k}_{\parallel}, \omega)\nonumber\\&~~~~\left[ {\hat{R}}_1(\VF{k}_{\parallel}, \omega)- {\hat{R}}_1^{\dagger}(\VF{k}_{\parallel}, \omega) \right]\hat{D}_{12}^{\dagger}(\VF{k}_{\parallel}, \omega)e^{-2\alpha d}\},
\label{eq:wideeq1to2}\end{align}for evanescent waves. The heat flux density from body 2 to body 1 can be obtained from Eqs.~\ref{eq:wideeq1to2prop} and \ref{eq:wideeq1to2} by exchanging the subscripts 1 and 2. We note that the derivation above does not assume reciprocity. The results are therefore generally applicable for either reciprocal or nonreciprocal systems. }
\section{\label{Constraint_sec}Constraint from the second law of thermodynamics}
\par{In this section, we show that Eqs.~\ref{eq:wideeq1to2prop} and \ref{eq:wideeq1to2} satisfy the second law of thermodynamics, by showing that the heat flux from body 1 to body 2 is balanced with the heat flux from body 2 to body 1, at each frequency $\omega$, and in-plane wavevector $\VF{k}_{\parallel}$, i.e.,
\begin{align}
 S_{1\to2}(\VF{k}_{\parallel},\omega)= S_{2\to1}(\VF{k}_{\parallel},\omega).
 \label{eq:Global_constraint}
\end{align}}\par{We start by providing a direct proof of Eq.~\ref{eq:Global_constraint} from Eqs.~\ref{eq:wideeq1to2prop} and \ref{eq:wideeq1to2}. For this purpose, we first state a few mathematical observations: We recall Eq.~\ref{Eq:Multiple_scattering} above as well as the analogous relation: \begin{align}
\hat{R}_{1}\hat{D}_{21} = \hat{D}_{12}\hat{R}_{1}   \label{eq:R1D21}.\end{align}By expanding $\hat{D}_{12}$, $\hat{D}_{21}$ in series, we note that
\begin{align}
\hat{D}_{12} &= \hat{I}+  \hat{R}_1\hat{D}_{21}\hat{R}_2e^{-j2k_{\rm z}d}
\label{simp1},\\
 \hat{D}_{21} &= \hat{I}+\hat{R}_2\hat{D}_{12}\hat{R}_1e^{-j2k_{\rm z}d}\label{simp2}.
\end{align}
\par{First, we consider the case of propagating waves. The heat flux $S_{1\to2} (\VF{k}_{\parallel},\omega)$ can be written as the sum of four terms, i.e.,}
\begin{align}
    S_{1\to2} (\VF{k}_{\parallel},\omega) =  {\cal Z}_1+{\cal Z}_2 + {\cal Z}_3 + {\cal Z}_4,
    \label{Eq:Prop_expand}
\end{align}where 
\begin{align}
    {\cal Z}_1 &=- \Tr[ \hat{R}_2^{\dagger} \hat{R}_2\hat{D}_{12} \hat{D}_{12}^{\dagger}]= -\Tr[  \hat{R}_2\hat{D}_{12} \hat{D}_{12}^{\dagger} \hat{R}_2^{\dagger}],\label{Z1_eq}\\
    {\cal Z}_2 &=-\Tr[  \hat{D}_{12} \hat{R}_1 \hat{R}_1^{\dagger}\hat{D}_{12}^{\dagger}]= -\Tr[  \hat{R}_1\hat{D}_{21} \hat{D}_{21}^{\dagger}\hat{R}_1^{\dagger} ].\label{Z2_eq}\end{align}\noindent{From Eqs. \ref{Z1_eq} and \ref{Z2_eq}, it can be shown that ${\cal Z}_1+{\cal Z}_2$ is symmetric with respect to exchanging the subscripts 1 and 2. Furthermore, for the last two terms in Eq.~\ref{Eq:Prop_expand}}, we can write:
    \begin{align}
    {\cal Z}_3 &=\Tr[\hat{D}_{12} \hat{D}_{12}^{\dagger}]\nonumber\\
   &= \Tr[\hat{I}+\hat{R}_1\hat{D}_{21} \hat{R}_2 (\hat{R}_1\hat{D}_{21} \hat{R}_2)^{\dagger}]\nonumber\\&~+\Tr\left[{\hat{R}_1\hat{D}_{21}\hat{R}_2e^{-j2k_{\rm z}d}}\right]+\Tr\left[{\hat{R}_1\hat{D}_{21}\hat{R}_2e^{-j2k_{\rm z}d}}{}\right]^{\dagger}\label{Z3_proof},\\
    {\cal Z}_4 &=\Tr[\hat{R}_2^{\dagger} \hat{R}_2\hat{D}_{12} \hat{R}_1 \hat{R}_1^{\dagger}\hat{D}_{12}^{\dagger}]= \Tr[ \hat{R}_2\hat{D}_{12} \hat{R}_1 (\hat{R}_2\hat{D}_{12} \hat{R}_1)^{\dagger}].\label{Eq_26_new}\end{align}
From Eqs.~\ref{Z3_proof}, \ref{Eq_26_new}, ${\cal Z}_3+{\cal Z}_4$ is also symmetric with respect to the subscript exchange between bodies 1 and 2. This can be seen by applying $\Tr\left[\hat{R}_1\hat{D}_{21}\hat{R}_2\right]$=$\Tr\left[\hat{R}_2\hat{D}_{12}\hat{R}_1\right]$ as derived from Eqs.~\ref{Eq:Multiple_scattering}, \ref{eq:R1D21}. This proves Eq.~\ref{eq:Global_constraint} for propagating waves.}
\par{We now consider the case of evanescent waves, and we write $S_{1\to2}(\VF{k}_{\parallel},\omega)$ as a sum of four terms (Eq.~\ref{Eq:Prop_expand}), where: 
\begin{align}
    {\cal Z}_1 &=\Tr\left[ \hat{R}_2^{\dagger}\hat{D}_{12}{\hat{R}}_1\hat{D}_{12}^{\dagger}\right]e^{-2\alpha d}\nonumber\\&= \Tr\left[\hat{D}_{12}{\hat{R}}_1(\hat{R}_2\hat{D}_{12})^{\dagger}\right]e^{-2\alpha d},\label{Eq_28}\\
    {\cal Z}_2 &=\Tr\left[ \hat{R}_2\hat{D}_{12}{\hat{R}}_1^{\dagger}\hat{D}_{12}^{\dagger}\right]e^{-2\alpha d}\nonumber\\&= \Tr\left[ \hat{D}_{21}\hat{R}_2(\hat{R}_1\hat{D}_{21})^{\dagger}\right]e^{-2\alpha d}.\label{Eq_29}\end{align}
From Eqs.~\ref{Eq_28}, \ref{Eq_29}, ${\cal Z}_1+{\cal Z}_2$ remains symmetric with respect to the exchange between bodies 1 and 2. Furthermore, 
 \begin{align}
    {\cal Z}_3 &=\Tr\left[- \hat{R}_2\hat{D}_{12}{\hat{R}}_1\hat{D}_{12}^{\dagger}\right]e^{-2\alpha d}=\Tr\left[- \hat{D}_{21}\hat{D}_{12}^{\dagger} + \hat{D}_{12}^{\dagger}\right]\label{Z_3_proof_evan},\\
    {\cal Z}_4 &=\Tr\left[- \hat{R}_2^{\dagger}\hat{D}_{12}{\hat{R}}_1^{\dagger}\hat{D}_{12}^{\dagger}\right]e^{-2\alpha d}=\Tr\left[-\hat{D}_{12}\hat{D}^{\dagger}_{21}+\hat{D}_{12}\right].
\end{align}
It can, therefore, be seen that ${\cal Z}_3+{\cal Z}_4$ also is symmetric with respect to exchanging bodies 1 and 2, using $\Tr\hat{D}_{12} = \Tr\hat{D}_{21}$ via Eqs.~\ref{simp1}, \ref{simp2}. This proves Eq.~\ref{eq:Global_constraint} for every frequency $\omega$ and in-plane wavevector $\VF{k}_{\parallel}$, of evanescent waves. Thus, we have proven here that Eqs.~\ref{eq:wideeq1to2prop} and \ref{eq:wideeq1to2} obey the second law of thermodynamics, even in the presence of nonreciprocity. Furthermore, Eq.~\ref{eq:Global_constraint} represents a general constraint on the heat transfer between two planar bodies, which must be satisfied in both reciprocal and non-reciprocal systems. }
\par
\begin{figure}[b]
 \includegraphics[scale = 0.3]{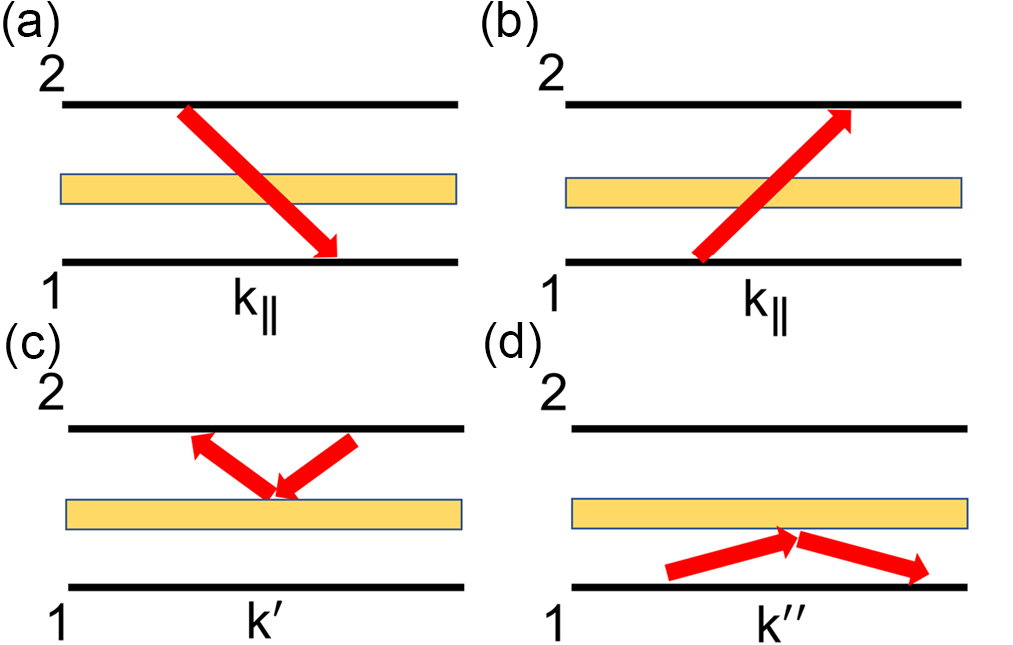}
\caption{Radiative heat transfer between bodies 1 and 2 in the presence of a lossless nonreciprocal filter placed in the vacuum gap, depicted with yellow. (a)-(b) The filter allows plane waves with a specific in-plane wavevector component $\VF{k}_{\parallel}$ to pass through. (c)-(d) The filter prohibits all other plane waves with in-plane wavevector components different from $\VF{k}_{\parallel}$ from passing through.}
\label{fig:fig3_yes}
\end{figure}

{We also note that the second law of thermodynamics requires the net energy exchange between bodies 1 and 2 to be zero, when $T_1=T_2 = T$, which can be written as: $S_{1\to2} = S_{2\to1}$. From Eqs.~\ref{S12_integral_first} and \ref{eq:Global_constraint}, it can be seen that Eq.~\ref{eq:Global_constraint} satisfies the second law of thermodynamics. Next, suppose we place in a lossless filter between the two bodies, as shown in Fig.~\ref{fig:fig3_yes}. The filter allows unity transmission from body 1 to 2 and vice versa, at $\omega$ and $\VF{k}_{\parallel}$ as shown in Figs. \ref{fig:fig3_yes}a and b, and completely reflects at all other frequencies and in-plane wavevectors as shown in Figs. \ref{fig:fig3_yes}c and d. In the presence of such a filter, when the two bodies are maintained at the same temperature $T$, we expect that the net energy transfer between the two bodies must be zero, thereby intuitively suggesting the constraint given by Eq.~\ref{eq:Global_constraint}.}

\section{\label{Sec:rec_const}Constraints from reciprocity}
\par{In this section, we show that in reciprocal systems, i.e. systems where both planar bodies consist of reciprocal materials, an additional constraint arises:  \begin{align}
    S_{1\to2}(\VF{k}_{\parallel},\omega) &= S_{2\to 1}( -\VF{k}_{\parallel},\omega)
    \label{eq:reciprocalkminusk}.
\end{align}To prove Eq.~\ref{eq:reciprocalkminusk}, we start by observing that the scattering matrix of a reciprocal system is symmetric [\onlinecite{haus_book}]. Hence, its reflectivity matrix satisfies: 
\begin{align}
    {\hat{R}}_{\rm i}(-\VF{k}_{\parallel}) &=  \hat{\sigma}_{\rm z}{\hat{R}}^T_{\rm i}(\VF{k}_{\parallel})\hat{\sigma}_{\rm z}\label{Ricorr_reciproca},
\end{align}where $\hat{\sigma}_{\rm z} = {\rm diag}[1, -1]$. Note that the matrix $\hat{\sigma}_{\mathrm z}$ appears because of the electric field components for the $p$-polarization acquire opposite signs for $\VF{k}_{\parallel}$ and $-\VF{k}_{\parallel}$. Henceforth, the argument $\omega$ is suppressed for brevity, since the derivation applies to each frequency separately. From Eq.~\ref{Ricorr_reciproca}, we have ${{\hat{{D}}}}_{12}(-\VF{k}_{\parallel})= \hat{\sigma}_{\rm z}{{\hat{{D}}}}^T_{21}(\VF{k}_{\parallel})\hat{\sigma}_{\rm z}$, as can be proved by inspecting $\hat{D}_{12}$ and $\hat{D}_{21}$ in the form of its series expansion. }
\par{Similar to the treatment in the previous section, we first consider propagating waves, for which: 
\begin{align}
 &~~~~S_{1\to2}( -\VF{k}_{\parallel})\nonumber\\
  &=\Tr\{\hat{\sigma}_{\rm z}\left[\hat{I} - \hat{R}_2^{*}(\VF{k}_{\parallel})\hat{R}_2^T(\VF{k}_{\parallel})\right]{\hat{D}}_{21}^T(\VF{k}_{\parallel})\nonumber\\ &~~~~\left[\hat{I} - {\hat{R}}^T_1(\VF{k}_{\parallel}){\hat{R}}_1^{
    *}(\VF{k}_{\parallel})\right] {\hat{D}}_{21}^{*}(\VF{k}_{\parallel}) \hat{\sigma}_{\rm z}\}\nonumber\\&=\Tr\{\left[\hat{I} - \hat{R}_1^{\dagger}(\VF{k}_{\parallel})  \hat{R}_1(\VF{k}_{\parallel})\right]{\hat{D}}_{21}(\VF{k}_{\parallel})\nonumber\\&~~~~ \left[\hat{I} - {\hat{R}}_2(\VF{k}_{\parallel}){\hat{R}}_2^{\dagger}(\VF{k}_{\parallel})\right] {\hat{D}}_{21}^{\dagger}(\VF{k}_{\parallel})\}\nonumber \\
    &=S_{2\to1}( \VF{k}_{\parallel}).
\end{align}Similarly, for evanescent waves, it holds: 
\begin{align}
 &~~~~S_{1\to2}( -\VF{k}_{\parallel})\nonumber\\
  &=\Tr\{\hat{\sigma}_{\rm z}\left[ \hat{R}_2^{*}(\VF{k}_{\parallel})-\hat{R}_2^T(\VF{k}_{\parallel})\right]{\hat{D}}_{21}^T(\VF{k}_{\parallel})\nonumber\\ &~~~~\left[ {\hat{R}}^T_1(\VF{k}_{\parallel})-{\hat{R}}_1^{*}(\VF{k}_{\parallel})\right] {\hat{D}}_{21}^{*}(\VF{k}_{\parallel}) \hat{\sigma}_{\rm z}e^{-2\alpha d}\}\nonumber\\&=\Tr\{\left[ \hat{R}_1^{\dagger}(\VF{k}_{\parallel})  -\hat{R}_1(\VF{k}_{\parallel})\right]{\hat{D}}_{21}(\VF{k}_{\parallel})\nonumber\\&~~~~ \left[ {\hat{R}}_2(\VF{k}_{\parallel})-{\hat{R}}_2^{
    \dagger}(\VF{k}_{\parallel})\right] {\hat{D}}_{21}^{\dagger}(\VF{k}_{\parallel}) e^{-2\alpha d}\}\nonumber\\
    &=S_{2\to1}(\VF{k}_{\parallel}).
\end{align}}\par{Hence, for any system consisting of two planar bodies, if Eq.~\ref{eq:reciprocalkminusk} is violated, then Eq.~\ref{Ricorr_reciproca} must also be violated for at least one of the bodies. Thus, at least one of the bodies must consist of a nonreciprocal material. Therefore, the violation of Eq.~\ref{eq:reciprocalkminusk} is a uniquely nonreciprocal effect in heat transfer between planar bodies. We can refer to the violation of Eq.~\ref{eq:reciprocalkminusk} as a nonreciprocal heat transfer effect. By contrast, however, satisfying Eq.~\ref{eq:reciprocalkminusk} does not guarantee that the underlying system is reciprocal. In fact, one can construct systems in which the reflectivity matrix of each individual body violates Eq.~\ref{Ricorr_reciproca}, and hence each individual body is nonreciprocal, but the resulting heat transfer ought to satisfy Eq.~\ref{eq:reciprocalkminusk}. We provide such an example in the numerical demonstration section below.}
\begin{figure*}
\includegraphics[scale = 0.3]{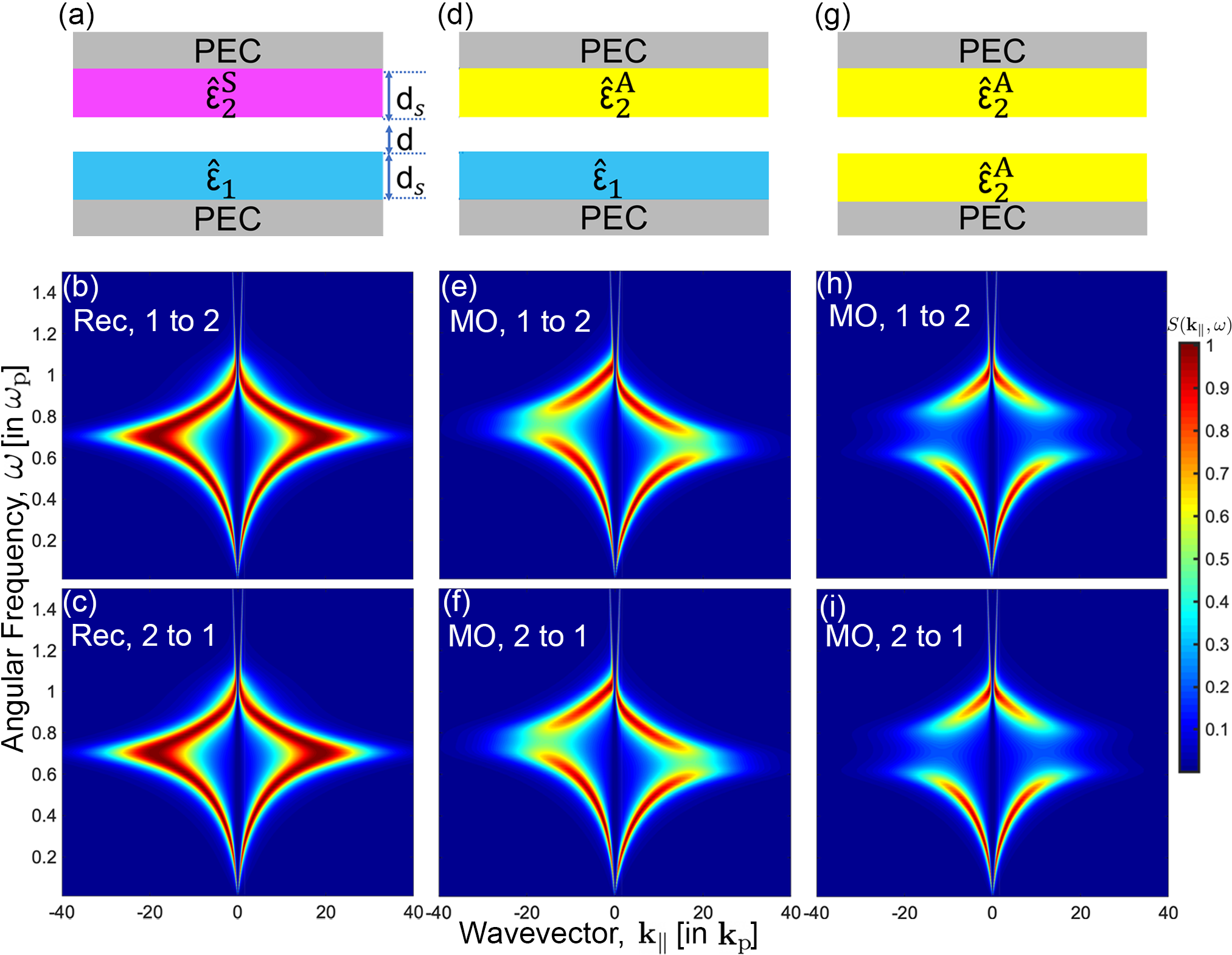} 
\caption{\label{fig:fig4_yes}Numerical simulations of heat transfer between two planar bodies. Body 1 is an isotropic plasmonic medium in panels (a)-(f), and a magneto-optical material with an asymmetric permittivity tensor ($\hat{\epsilon}_2^{\rm A}$ in Eq.~\ref{subeq:asyms}) in panels (g)-(i). Body 2 is an anisotropic material with a symmetric dielectric permittivity tensor ($\hat{\epsilon}_2^{\rm S}$ in Eq.~\ref{subeq:1}) in panels (a)-(c), and a magneto-optical material with $\hat{\epsilon}_2^{\rm A}$ (Eq.~\ref{subeq:asyms}) in panels (d)-(i). The thickness of the slabs is set to $d_{\rm S}=\lambda_{\rm p}$, whereas the vacuum gap has width of $d=0.1\lambda_{\rm p}$, where $\lambda_{\rm p}$ is bulk plasma wavelength. Panels (b), (e), (h) represent heat flux density from body 1 to 2, i.e., $S_{1\to 2}$, panels (c), (f), (i) represent heat flux density from body 2 to 1, i.e., $S_{2\to 1}$. Panels (e), (f) demonstrate the violation of reciprocity constraint for heat transfer, i.e., Eq.~\ref{eq:reciprocalkminusk}, whereas panels (b), (c) and (h), (i) satisfy Eq.~\ref{eq:reciprocalkminusk}. All results agree with the general thermodynamic constraint of Eq.~\ref{eq:Global_constraint}. }
\end{figure*}
\section{\label{numerical_sec}Numerical Demonstrations}
In this section, we demonstrate the findings of the previous sections by performing numerical calculations of heat transfer between two planar slabs. First, we consider a reciprocal system consisting of two planar slabs of reciprocal materials, as shown in Fig.~\ref{fig:fig4_yes}a. We set body 1 to have an isotropic dielectric permittivity, $\hat{\epsilon}_1 = \epsilon_{\rm p}$, where $\epsilon_{\rm p}$ is the dielectric function of a plasmonic metal that takes the Drude model form $\epsilon_{\rm p} = 1-\frac{\omega_{\rm p}^2}{\omega(\omega+j/\tau)}$. $\omega_{\rm p}$ is the plasma frequency, and $1/\tau = 0.1\omega_{\rm p}$ characterizes the plasmonic scattering rate. We choose the permittivity of the second body to be anisotropic and have the form:
\begin{align}
\hat{\epsilon}_2^{\rm S}(\omega) &= \left[\begin{array}{ccc}
  \epsilon_{\rm p} &  0& 0 \\
  0 &  \epsilon_{\rm d} & \epsilon_{\rm f} \\
   0 &   \epsilon_{\rm f}&  \epsilon_{\rm d}
\end{array}\right]\label{subeq:1},\end{align}with $\epsilon_{\rm d} = 1-\frac{\omega_{\rm p}^2(1+j\frac{1}{\tau\omega})}{(\omega+j/\tau)^2-\omega_{\rm B}^2}$, $\epsilon_{\rm f} = -\frac{\omega_{\rm B}}{\omega}\frac{\omega_{\rm p}^2}{(\omega+j/\tau)^2-\omega_{\rm B}^2}$. Here, $\omega_{\rm B}$ is chosen to be $\omega_{\rm B}=0.2\omega_{\rm p}$. The permittivity form of Eq.~\ref{subeq:1} is chosen to facilitate the comparison to the nonreciprocal case as we will show below. This form resembles the permittivity of a magnetized plasma, except that the permittivity tensor is symmetric and hence the material is reciprocal. We define the plasma wavelength $\lambda_{\rm p}=2\pi c/\omega_{\rm p}$, and set the thickness of both slabs as $d_{\rm s}=\lambda_{\rm p}$. The width of the vacuum gap between two slabs is set to be $d=\lambda_{\rm p}/10$, for probing near-field effects. To remove the effect of any bulk free-space propagating modes that occur at high frequencies, we place a perfect electric conductor (PEC) on the back-side of both slabs.

Based on the formalism described in Section~\ref{setup_sec}, we compute the radiative heat transfer between these two bodies, for the in-plane wavevector $\VF{k}_{\parallel}$ along the $y$-direction, as shown in Figs.~\ref{fig:fig4_yes}b and c, corresponding to $S_{1\to 2}(\VF{k}_{\parallel},\omega)$ and $S_{2\to1}(\VF{k}_{\parallel},\omega)$, respectively. We see that $S_{1\to 2}(\VF{k}_{\parallel},\omega) =S_{1\to 2}(-\VF{k}_{\parallel},\omega)$, as expected for a reciprocal system (Eq.~\ref{eq:reciprocalkminusk}). Furthermore, for each frequency and in-plane wavevector, we have $S_{1 \to 2} (\VF{k}_{\parallel}, \omega) = S_{2\to 1}(\VF{k}_{\parallel}, \omega)$, confirming the validity of Eq.~\ref{eq:Global_constraint}. We emphasize that these constraints are satisfied despite the lack of any mirror symmetry in the system. Bodies 1 and 2 are made of different materials, and moreover the permittivity of body 2 is chosen to be anisotropic such that different $\VF{k}_{\parallel}$ are not equivalent.  

Next, we study a nonreciprocal case, where Eq.~\ref{eq:reciprocalkminusk} is violated. We consider a system where body 1 remains the same as in the previous example, however body 2 has a permittivity tensor of the form (Fig. \ref{fig:fig4_yes}d): 
\begin{align}\hat{\epsilon}_2^{\rm A}(\omega) &= \left[\begin{array}{ccc}
  \epsilon_{\rm p} &  0& 0 \\
  0 &  \epsilon_{\rm d} & j\epsilon_{\rm f} \\
   0 &   -j\epsilon_{\rm f} &  \epsilon_{\rm d}
\end{array}\right]\label{subeq:asyms}.
\end{align}This permittivity tensor is asymmetric, thus breaking Lorentz reciprocity. The heat transfer spectra for this system are shown in Figs.~\ref{fig:fig4_yes}e and f. We see, here, $S_{1\to2}(\VF{k}_{\parallel},\omega)\neq S_{1\to2}(-\VF{k}_{\parallel},\omega)$. Hence, based on the discussions in Section~\ref{Sec:rec_const}, we have shown that this system can achieve nonreciprocal near-field heat transfer in a planar geometry. Furthermore, by comparing Figs.~\ref{fig:fig4_yes}e and f, we see that $S_{1\to2}(\VF{k}_{\parallel},\omega)=S_{2\to1}(\VF{k}_{\parallel},\omega)$ for all frequencies $\omega$ and in-plane wavevectors $\VF{k}_{\parallel}$, in consistency with Eq.~\ref{eq:Global_constraint}. This numerical example demonstrates that Eq.~\ref{eq:Global_constraint} is satisfied despite the lack of both reciprocity and mirror symmetry in the considered system. 

Finally, we consider a nonreciprocal case, which demonstrates that even in the presence of nonreciprocity, Eq.~\ref{eq:reciprocalkminusk} can hold. Hence, we set both bodies to have the permittivity tensor of Eq.~\ref{subeq:asyms}. In this case, the signature of nonreciprocity by each individual body is cancelled out via their combination. Consequently we have $S_{1\to2}(\VF{k}_{\parallel},\omega)=S_{1\to2}(-\VF{k}_{\parallel},\omega)$ despite the fact that the response for each individual body is nonreciprocal. Furthermore, similar to the previous cases, the constraint $S_{1\to2}(\VF{k}_{\parallel},\omega)=S_{2\to1}(\VF{k}_{\parallel},\omega)$ is preserved since Fig.~\ref{fig:fig4_yes}h is identical to Fig.~\ref{fig:fig4_yes}i, once again verifying the general thermodynamic constraint Eq.~\ref{eq:Global_constraint}.
\par{To conclude this section, we provided a numerical demonstration showing that Eq.~\ref{eq:Global_constraint} holds both in reciprocal (Fig.~\ref{fig:fig4_yes}a) and nonreciprocal systems (Figs.~\ref{fig:fig4_yes}d, g). Furthermore, we showed that the generalized detailed balance (Eq.~\ref{eq:reciprocalkminusk}) holds for reciprocal systems, and the violation of Eq.~\ref{eq:reciprocalkminusk} can only be achieved in nonreciprocal systems. Finally, we demonstrated a case where Eq.~\ref{eq:reciprocalkminusk} holds even when the underlying bodies are nonreciprocal. 
\section{\label{conclude_sec}Conclusion}
We presented a formalism for computing radiative heat transfer between two planar bodies. This formalism is applicable for both reciprocal and nonreciprocal systems. We introduced a constraint imposed by the second law of thermodynamics and reciprocity, that holds for every in-plane wavevector and frequency. Therefore, our formalism identifies the unique signature of nonreciprocity in heat transfer in two-body planar systems.}
\begin{acknowledgments}
Discussions with Dr. Linxiao Zhu and Dr. Weiliang Jin are gratefully acknowledged.  This work is supported by an U. S. Army Research Office (ARO) MURI Grant W911NF-19-1-0279. G.T.P. acknowledges the TomKat Postdoctoral Fellowship in Sustainable Energy at Stanford University. Z.Z. and S. B. acknowledge the support of a Stanford Graduate Fellowship. 
\end{acknowledgments}
\appendix
\setcounter{secnumdepth}{0}
\section{Appendix}
\par{In this Appendix, we provide a proof of Eqs.~\ref{eq:EEcorre_prop1} and~\ref{eq:EEcorre_evan1} presented in the main text, regarding the fields emitted by a single planar body, based on the generalized reciprocity theorem. Eqs.~\ref{eq:EEcorre_prop1} and~\ref{eq:EEcorre_evan1} have been previously proven in Refs.~\onlinecite{PhysRevA_correlation, PhysRevA.80.042102, Messina2011Scattering, doi:10.1063/1.4883243} by using the second fluctuation-dissipation theorem, which directly relates the field emission to the imaginary part of the Green's function. The proof of Refs.~\onlinecite{PhysRevA_correlation, PhysRevA.80.042102, Messina2011Scattering, doi:10.1063/1.4883243} applies to both reciprocal and nonreciprocal systems. Here, we provide an alternative proof based entirely on the current correlation. This proof facilitates the understanding of the differences between reciprocal and nonreciprocal systems.}

\par{We first summarize a few relations in electromagnetics that we will use below, applicable to reciprocal as well as nonreciprocal systems. For simplicity, for both theorems discussed below, we consider a non-magnetic ($\mu=\mu_0$) system, described by relative dielectric permittivity $\hat{\epsilon}$.}

\par{\emph{Theorem 1. Conservation of energy.} Let us consider a system that has two steady-state solutions as described by electric fields $\VF{E}_{\rm \sigma}$ and $\VF{E}_{\rm \mu}$. At each frequency $\omega$, we have:
 \begin{align}
      &~~~~\int\dif{V}  \omega\epsilon_0\VF{E}_{\rm \sigma}^{*}\cdot\left[\frac{\hat{\epsilon}-\hat{\epsilon}^{\dagger}}{2j}\right]\VF{E}_{\rm \mu} \nonumber\\&=-\frac{1}{2}\int\dif{S}\hat{\VF{n}}\cdot(\VF{E}_{\rm \sigma}^*\times\VF{H}_{\rm \mu} + \VF{E}_{\rm \mu}\times\VF{H}_{\rm \sigma}^*).
      \label{general_poynting_integral}
  \end{align} }\noindent{The proof of Eq.~\ref{general_poynting_integral} can be found in Ref. [\onlinecite{Kong_poynting_new}]. In the geometry of Fig.~\ref{fig:fig2_yes}a, where body $1$ occupies the half space $z<0$, suppose that the fields $\VF{E}_{\sigma}(z)e^{-j\VF{k}_{\parallel}\cdot\VF{r}_{\parallel}}$, $\VF{H}_{\sigma}(z)e^{-j\VF{k}_{\parallel}\cdot\VF{r}_{\parallel}}$ and $\VF{E}_{\mu}(z)e^{-j\VF{k}_{\parallel}\cdot\VF{r}_{\parallel}}$, $\VF{H}_{\mu}(z)e^{-j\VF{k}_{\parallel}\cdot\VF{r}_{\parallel}}$ are solutions to Maxwell’s equations at the in-plane wavevector $\VF{k}_{\parallel}$. Then, Eq.~\ref{general_poynting_integral} becomes:
  \begin{align}
      &~~~~\int_{-\infty}^0\dif{z}  \omega\epsilon_0\VF{E}_{\rm \sigma}^{*}\cdot\left[\frac{\hat{\epsilon}-\hat{\epsilon}^{\dagger}}{2j}\right]\VF{E}_{\rm \mu}\nonumber\\&=-\frac{1}{2}(\VF{E}_{\rm \sigma}^*\times\VF{H}_{\rm \mu} + \VF{E}_{\rm \mu}\times\VF{H}_{\rm \sigma}^*)\cdot\hat{\VF{z}}\mid_{z = 0^+}.
 \label{Poynting_z_decomp}
  \end{align}}\par{\emph{Theorem 2. Generalized reciprocity theorem.} For nonmagnetic systems, the electric field $\VF{E}$ can be obtained by solving the equation $\nabla\times\nabla\times\VF{E} - k_0^2\hat{\epsilon}\VF{E} =-j\omega\mu_0\VF{J}$, where $k_0 = \omega/c$ and $\VF{J}$ is the current density. The Green’s function for this system is defined by:
\begin{align}
    \nabla\times\nabla\times G - k_0^2\hat{\epsilon} G &=\mathbb{I}\delta(\VF{r}-\VF{r}'),
\end{align}where $\mathbb{I}$ is the identity dyad. Generalized reciprocity states:
\begin{align}
    \Tilde{G}(\VF{r}', \VF{r})  &= {G}^T(\VF{r}, \VF{r}')\label{GfuncRRp},\end{align}where $\tilde{G}$ is the Green’s function of its complementary system described by a permittivity of $\hat{\epsilon}^T$. For reciprocal systems, $\hat{\epsilon} = \hat{\epsilon}^T$, and hence $G(\VF{r}',\VF{r}) = G^T(\VF{r},\VF{r}')$. The proof of Eq.~\ref{GfuncRRp} closely parallels the standard proof in reciprocal systems, and can be found in Ref. [\onlinecite{Kong_reciprocity_new}].}

\par{Next, we consider the field emission in Eqs.~\ref{eq:EEcorre_prop1} and~\ref{eq:EEcorre_evan1}. For the system shown in Fig. \ref{fig:fig2_yes}, by taking the Fourier transform of $\tilde{G}$ and $G$, we obtain:
\begin{align}
    \Tilde{G}(\VF{k}_{\parallel}, z', z_0)  &= {G}^T(-\VF{k}_{\parallel},z_0, z' ).\label{Eq:Grelation_kx_ky}
\end{align}}\par{For the system shown in Fig.~\ref{fig:fig2_yes}a, we calculate the electric field emission of $\VF{E}_{1}(\VF{k}_{\parallel}, z, \omega)$ near the surface $z=0^+\equiv z_0$. Using the relation $\VF{E}(\VF{r}) =-j\omega\mu_0\int\dif{\VF{r}}'G(\VF{r}, \VF{r}')\cdot \VF{J}(\VF{r}')$, as well as the fluctuation dissipation theorem [\onlinecite{1989psr4bookFDT}], which takes the following form:
\begin{align}
       &~~~~ \left<\VF{J}(\VF{k}_{\parallel}, \omega, z)\VF{J}^{\dagger}(\VF{k}', \omega, z')\right>\nonumber\\ &= (2\pi)^2\frac{4}{\pi}\omega\epsilon_0\Theta(\omega,T)\left[\frac{\hat{{\epsilon}} - \hat{{\epsilon}}^{\dagger}}{2j}\right]\delta(\VF{k}_{\parallel} - \VF{k}'_{\parallel})\delta(z-z'),
   \end{align} \noindent{we obtain the electric field emission in Eqs.~\ref{eq:EEcorre_prop1} and~\ref{eq:EEcorre_evan1} in the following form:}
\begin{align}
&~~~~\left<\VF{E}_1\VF{E}_1^{\dagger}\right>\nonumber\\&= (\omega\mu_0)^2\int\dif{z'}\int\dif{z''}G(\VF{k}_{\parallel}, z_0, z')\nonumber\\&~~~~\left<\VF{J}(\VF{k}_{\parallel}, \omega, z') \VF{J}^{\dagger}(\VF{k}_{\parallel},\omega, z'')\right>G^{ \dagger}(\VF{k}_{\parallel},z_0, z'')\nonumber\\
    &= (2\pi)^2\frac{4}{\pi}\Theta(\omega,T)(\omega\mu_0)^2\nonumber\\&~~~~\int_{-\infty}^0\dif{z'}\omega\epsilon_0G(\VF{k}_{\parallel},z_0, z')\left[\frac{{\hat{\epsilon}} - {\hat{\epsilon}}^{\dagger}}{2j}\right]G^{\dagger}(\VF{k}_{\parallel},z_0, z')\nonumber\\
    &=(2\pi)^2\frac{4}{\pi}\Theta(\omega,T)(\omega\mu_0)^2\nonumber\\&~~~~ \left[\int_{-\infty}^0\dif{z}' \omega\epsilon_0 \Tilde{G}^{\dagger}(-\VF{k}_{\parallel}, z',z_0)\left[\frac{\Tilde{\epsilon} - \Tilde{\epsilon} ^{\dagger}}{2j}\right]\Tilde{G}(-\VF{k}_{\parallel},z', z_0 )\right]^T,
    \label{F_Gfunctions}
\end{align}\noindent{where in the last step we have used generalized reciprocity as described in Eq. \ref{Eq:Grelation_kx_ky}. Here, the electric field $\VF{E}_1$ is expressed as $\VF{E}_1 = E_{\rm 1s} {\VF{s}}+E_{\rm 1p} {\VF{p}}_+$, where $\VF{s}$ and $\VF{p}_+$ are the polarization unit vectors for the $s-$ and $p-$polarized outgoing waves, in consistency with the definition of reflectivity matrix in Eq.~\ref{reflec_mat}. For subsequent use, we also define $\VF{p}_-$ as the unit polarization vector for the incoming $p-$polarized wave. } }
\par{The Green's function, $\tilde{G}(-\VF{k}_{\parallel},z',z_0)$ for the complementary system $\tilde{\epsilon} = \hat{\epsilon}^T$, takes the dyadic form
   $ \Tilde{G}(-\VF{k}_{\parallel},z', z_0 )=\frac{1}{-j\omega\mu_0}[ \Tilde{\VF{E}}_{\rm s}\tilde{\VF{s}}+ \Tilde{\VF{E}}_{\rm p}\tilde{\VF{p}}_-]$.
Here, $\tilde{\VF{E}}_{\rm s,p}$ stand for the electric field at $z'$, generated from ${{s}}-$ or ${{p}}-$polarized current source at $z_0$ with unit amplitude in the complementary system. We define the polarization basis vectors $\tilde{\VF{s}}$, $\tilde{\VF{p}}_+$, $\tilde{\VF{p}}_-$ equivalently to $\VF{s}$, $\VF{p}_+$, $\VF{p}_-$, but for the in-plane wavevector $-{\VF{k}_{\parallel}}$. The polarization basis vectors are connected via $\tilde{\VF{s}} =-{\VF{s}}$, $\tilde{\VF{p}}_+ = {\VF{p}}_-$, $\tilde{\VF{p}}_- = {\VF{p}}_+$. In terms of the polarization basis $\tilde{\VF{s}}$ and $\tilde{\VF{p}}_-$, Eq.~\ref{F_Gfunctions} is expanded into four terms:\begin{align} \left<\VF{E}_1\VF{E}_1^{\dagger}\right> &=  (2\pi)^2\frac{4}{\pi}\Theta(\omega,T)\nonumber\\&~~~~ (F_{\rm ss}\tilde{\VF{s}}\tilde{\VF{s}}^{\dagger}+ F_{\rm sp}\tilde{\VF{p}}_-{\tilde{\VF{s}}}^{\dagger}+ F_{\rm ps}\tilde{\VF{s}}\tilde{\VF{p}}_-^{\dagger}+ F_{\rm pp}\tilde{\VF{p}}_-\tilde{\VF{p}}_-^{\dagger}).
\label{EEcorrelation_2vec}
\end{align} where
  \begin{align}
       F_{\rm \sigma\mu}  &= \int_{-\infty}^0\dif{z}'  \omega\epsilon_0\tilde{\VF{E}}_{\rm \sigma}^{*}\cdot\left[\frac{\hat{\epsilon}-\hat{\epsilon}^{\dagger}}{2j}\right]\tilde{\VF{E}}_{\rm \mu}\nonumber\\&=-\frac{1}{2}(\tilde{\VF{E}}_{\rm \sigma}^*\times\tilde{\VF{H}}_{\rm \mu} + \tilde{\VF{E}}_{\rm \mu}\times\tilde{\VF{H}}_{\rm \sigma}^*)\cdot\hat{\VF{z}}\mid_{z = z_0}
       \label{Eq:PoyntingUsingG}.
  \end{align}On the other hand, $\tilde{\VF{E}}_{\rm s} = -\frac{\omega\mu_0}{2k_{\rm z}}[(1+\tilde{R}^{\rm ss}_1)\tilde{\VF{s}} + \Tilde{R}^{\rm ps}_1\tilde{\VF{p}}_+]$, $\tilde{\VF{E}}_{\rm p} = -\frac{\omega\mu_0}{2k_{\rm z}}[\Tilde{R}_1^{\rm sp}\tilde{\VF{s}}+\tilde{\VF{p}}_-+\tilde{R}_1^{\rm pp}\tilde{\VF{p}}_+] $ are the electric fields near the surface $z =z_0\equiv 0^+$. The magnetic fields can be computed to be, $\VF{H}_{\rm s} = -\frac{k_0}{2k_{\rm z}}\left[-\tilde{\VF{p}}_--\tilde{R}^{\rm ss}_1\tilde{\VF{p}}_+ + \Tilde{R}^{\rm ps}_1\tilde{\VF{s}}\right]$ and $\VF{H}_{\rm p} = -\frac{k_0}{2k_{\rm z}}\left[-\Tilde{R}_1^{\rm sp}\tilde{\VF{p}}_++\tilde{\VF{s}}+\tilde{R}_1^{\rm pp}\tilde{\VF{s}}\right]$. By evaluating the right hand side of Eq.~\ref{Eq:PoyntingUsingG}, we obtain the following relation for propagating waves: 
\begin{align}
 4F/Z&= \left({1-|\tilde{R}_1^{\rm ss}|^2}{-|\tilde{R}_1^{\rm ps}|^2}\right){\VF{s}}{\VF{s}}^{\dagger}\nonumber\\&~~~~+\left( {\tilde{R}_1^{\rm sp*}\tilde{R}_1^{\rm ss}}+{\tilde{R}_1^{\rm ps}\tilde{R}_1^{\rm pp*}}\right){\VF{{s}}}{\VF{{p}}}_+^{\dagger}\nonumber\\& ~~~~+  \left( {\tilde{R}_1^{\rm sp}\tilde{R}_1^{\rm ss*}}+{\tilde{R}_1^{\rm ps*}\tilde{R}_1^{\rm pp}}\right){{\VF{p}}}_+{\VF{{s}}}^{\dagger}\nonumber\\&~~~~+  \left({1-|\tilde{R}_1^{\rm pp}|^2}{-|\tilde{R}_1^{\rm sp}|^2}\right){\VF{{p}}}_+{\VF{{p}}}_+^{\dagger},\end{align} \noindent{whereas for evanescent waves, we have: }
 \begin{align} 4 F/Z&= \left({\tilde{R}_1^{\rm ss}-\tilde{R}_1^{\rm ss*}}\right){\VF{s}}{\VF{s}}^{\dagger}\nonumber\\&~~~~- \left({\tilde{R}_1^{\rm sp*}}- {\tilde{R}_1^{\rm ps}}\right){\VF{{s}}}{\VF{{p}}}_+^{\dagger}\nonumber\\&~~~~- \left( {\tilde{R}_1^{\rm sp}} - {\tilde{R}_1^{\rm ps*}}\right){{\VF{p}}}_+{\VF{{s}}}^{\dagger}\nonumber\\&~~~~+ \left(\tilde{R}_1^{\rm pp}-{\tilde{R}_1^{\rm pp*}}\right){\VF{{p}}}_+{\VF{{p}}}_+^{\dagger}.
\end{align} Here, the reflectivity matrix $\tilde{R}_1$ corresponds to the complementary system with $\tilde{\epsilon}$ at the in-plane vector $-\VF{k}_{\parallel}$, in the form of Eq.~\ref{reflec_mat}. We also note that $Z = {\rm diag}[Z_{\rm s},Z_{\rm p}]$,  $Z_{\rm s} = Z_{\rm p} = Z_0\frac{k_0}{k_{\rm z}}$, and $Z_0=\sqrt{\frac{\mu_0}{\epsilon_0}}$ is the impedance
in vacuum. The reflectivity matrix $\hat{R}_1$ of the original system with $\hat{\epsilon}$ at $\VF{k}_{\parallel}$ is related to $\tilde{R}_1$ in the complementary system with $\tilde{\epsilon}$ at $-\VF{k}_{\parallel}$ via $\tilde{R}_1(-\VF{k}_{\parallel})= \hat{\sigma}_{\rm z}\hat{R}_1^T(\VF{k}_{\parallel})\hat{\sigma}_{\rm z}$, as can be proved via symmetry by rotating the coordinate axes to transform the dielectric permittivity tensor [\onlinecite{haus_book}]. Also, by recalling that $\hat{R}_1 = {R}_1^{\rm ss} {\VF{s}}{\VF{s}}+{R}_1^{\rm sp} {\VF{s}}{\VF{p}}_++{R}_1^{\rm ps} {\VF{p}}_-{\VF{s}}+{R}_1^{\rm pp} {\VF{p}}_-{\VF{p}}_+$, we express the field emission in Eq.~\ref{EEcorrelation_2vec} as:
  \begin{align}
 \left<\VF{E}_1\VF{E}_1^{\dagger}\right> &= (2\pi)^2\frac{Z}{\pi}\Theta(\omega,T_1)\nonumber\\&~~~~\begin{cases}
 \left[\hat{I}-{\hat{R}}_1{\hat{R}}_1^{\dagger}\right],~~~~~\text{propagating waves} \\
\left[{\hat{R}}_1-{\hat{R}}_1^{\dagger}\right],~~~~~~\text{ evanescent waves} 
\end{cases}
  \end{align}\noindent{We note that the derivation above does not use reciprocity, and hence is applicable to both reciprocal and nonreciprocal systems.}}


%
\end{document}